# Spatial and Spatio-Temporal Multidimensional Data Modelling: A Survey

Saida Aissi[1], Mohamed Salah Gouider[2],

Bestmod Laboratory. University of Tunis, High Institute of Management of Tunis, Tunisia

saida.aissi@yahoo.fr[1] @ ms;gouider[2]@yahoo.fr

ABSTRACT— **Data warehouse store and provide access to large volume of historical data supporting the strategic decisions of organisations. Data warehouse is based on a multidimensional model which allow to express user's needs for supporting the decision making process. Since it is estimated that 80% of data used for decision making has a spatial or location component [1, 2], spatial data have been widely integrated in Data Warehouses and in OLAP systems. Extending a multidimensional data model by the inclusion of spatial data provides a concise and organised spatial datawarehouse representation.**
**This paper aims to provide a comprehensive review of litterature on developed and suggested spatial and spatio-temporel multidimensional models. A benchmarking study of the proposed models is presented. Several evaluation criterias are used to identify the existence of trends as well as potential needs for further investigations.**

*Keywords*— **Dtawarehouse, Spatial data, Multidimensionnal modelling, Temporal data**

## I. INTRODUCTION

A data warehouse is defined as a collection of subject-oriented, integrated, non-volatile and time-variant data supporting management's decisions-making process [1].The conventional data warehouses are designed based on a multidimensional view of data. It is based on a fact table representing the subject orientation and the focus of analysis. The fact table contains usually numeric data called measures representing analysis needs in a quantified form. These measures are explored from different perspectives through dimensions.

With the development of computerized data collection tools like satellite telemetry systems, and medical imaging, a huge amount of spatial data have became a major component in the decisions-making process. In fact, it is estimated that 80% of data stored in databases has a spatial or location component [1, 2, 4]. Developing efficient methods for analysing and understanding of such a huge amount of spatial data becomes and imminent task. Developing spatial data warehouses was the prefect solution allowing online spatial data analysis and spatial data mining.

This paper reviews in the first section the diffe concepts that have emerged. The second section presents rent types of spatial and spatio-temporal multidimensional models that have been proposed as well as new a comparative study between the different proposed models as well as new line of research susceptible to be more exploited.

## II. SURVEY ON SPATIAL AND SPATIO-TEMPORAL MUTLIDIMENSIONNA MODELS

Despite the relatively short history of spatial and spatio-temporal modelling, a substantial number of models have been presented. In this section, we examines most of spatial and spatio-temporal multidimensional models proposed in the literature. Each model is discussed and evaluated relatively to the other models and a comparative study between the different models is realised.

### A. The model of Stefanovic and al

Stefanovic and al [5] proposes a spatial Data warehouse model presented in fig 1. This model consists of both spatial and non spatial dimensions and measures. A spatial measure is defined as a measure "which contains a collection of pointers to spatial objects". In this context, a star model for the analysis of weather is proposed. This model contains three measures and four dimensions: dimension precipitation, dimension region_name, dimension time and dimension temperature. Notice that the dimension region_name at the primitive level is a spatial object name representing the corresponding spatial region on the map. Of the three measures, the measure region_map is the only spatial measure; it contains a collection of spatial pointers pointing to the corresponding regions. The measures area and count are numerical measures. The measure area represents the sum of the total areas of the corresponding spatial objects. The measure count represents the total number of base regions. Such a configuration serves two purposes: to provide measures ( area and count) to users for exploration and analysis and to provide a link ( region_map) between portions of the hypercube and their spatial representations. According to the researchers, this link is required to dynamically perform Roll-UPs and Drill-Downs operations as it identifies spatial objects that have to be dynamically merged for representations purposes.

In this work, researchers propose three methods for computing spatial measures in spatial data cube construction. The first method is based on collecting and storing corresponding spatial objects pointers without performing precomputation of spatial measures in a spatial data cube. This solution can be implemented by storing, in the cube cell, a pointer to a collection of spatial objects pointers. The second method is based on computing and storing same rough approximations/ estimations of the spatial measures in a spatial data cube. Such a recomputed result is as small as a nonspatial measure and can be presented quickly to users. The third method is based on precomputing selectively some spatial measures The selection of the measures to be precomputed can be performed at the cuboid level. Example: either precompute and store each set of mergeable spatial region for each cell of a selected cuboid, or precompute none if the cuboid is non selected.

Contrary to other models, the present model supports the symmetric treatment of the dimensions and the measures. However, To propose their model of warehouse of data, the researchers suppose that the data arise from homogeneous sources which is a little realistic hypothesis. Otherwise, The problem of update of the data warehouse to guarantee the consistency of the data is not taken into account.

The use of estimations values for the calculation of the spatial measures is useful for results asking for general estimates. However, concerning results asking for more details, the application has to load the other results.

### B. The model of Stefanovic and al

Through the MultiDimER model, Malinowski and Zimany [6, 7, 8] propose an extension of a conceptual multidimensional model. They introduce spatial measures, spatial dimensions, spatial hierarchies as well as spatial topological relationships and topological operators in the model.

In order to introduce spatiality in the dimensions, researchers use the spatial data types developed in the MADS model. A spatial dimension is defined as a dimension that includes at least one spatial hierarchy. A spatial hierarchy is a hierarchy that includes at least one spatial level and a spatial level is defined as a level for which the application needs to keep some spatial characteristics ( eg; the geometry represented by spatial data types like point, area, collection of point or a collection of area). In the MultiDimER model, spatial levels are related to each other using topological relationships such as contain, intersects, equals, etc.

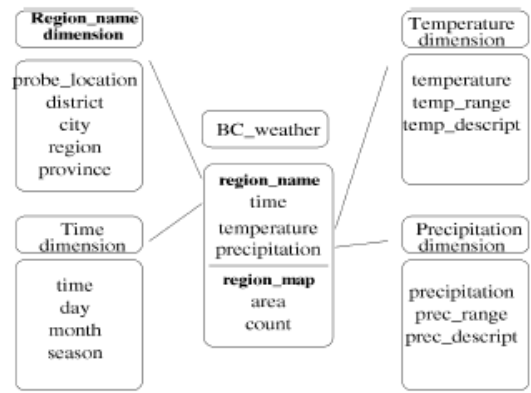

Figure 1: A star model for a spatial weather datawarehouse [5]

A model for the analysis of clients buying behaviour is presented. It contains a spatial dimension (dimension Client), a thematic dimension (dimension Product) and a temporal dimension (dimension Time). The spatial dimension is composed of spatial level of different kind: point for Store, area for City and a collection of area for State.

The analysis of clients buying behaviour is realised by using selection predicates (operations Slice and Dice) and by choosing the aggregation level (Roll-*up* and *Drill-down* operations).

In order to introduce spatial measure in the classical multidimensional model. Researchers distinguish two types of spatial measure. In fact, a spatial measure is defined as a measure that either is represented by geometry and uses a spatial function for aggregation along hierarchies, or represents a numerical value that is calculated using spatial or topological operators. A multidimensional model with spatial measure was proposed in fig 2.

The model is used for analysing locations of accidents tacking into account the different insurance categories. It includes a spatial geometric measure (the measure location) representing the location of an accident. A spatial aggregation function named geometric union is defined. So, when the user rolls-up to the Insurance Category level, the locations corresponding to different
categories will be aggregated and represented as a set of points.

In order to introduce the second type of spatial measure which is calculated using spatial or topological operators. A multidimensional model for analysing the closeness of saptial objects is proposed. This model includes a spatial holistic measure (*Min.distance*) that calculates the distance from states to highway sections using spatial operators.

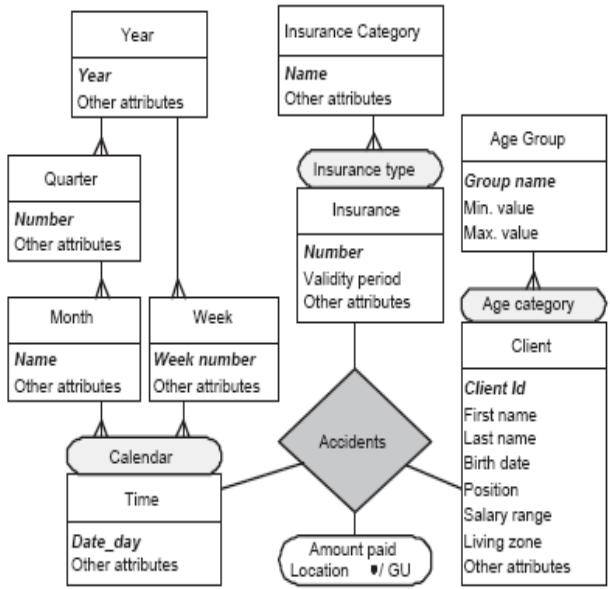

Figure 2 : A multidimensionnal model with a spatial measure (localisation) [6]

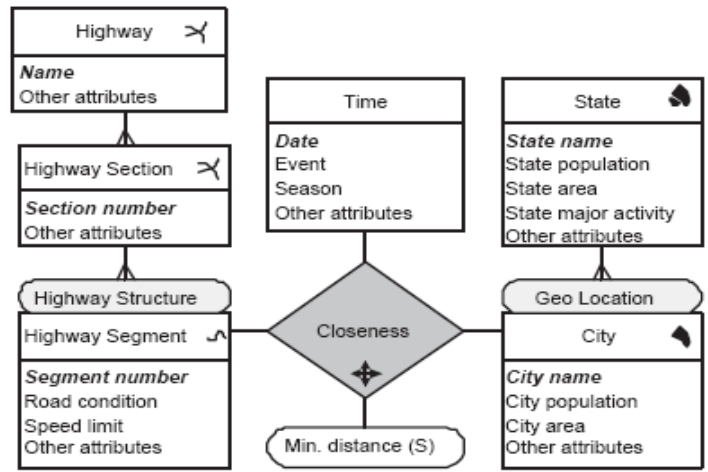

Figure 3: A multidimensionnal model with a spatial computed numéric measure [6]

Finally, a spatial fact relationship is defined as a fact relationship that requires a spatial join between two or more dimensions. Malinowski and Zimany propose to include a spatial predicate symbol in the fact relationship (fig 3). They propose a model for the analysis of highway maintenance cost in which the fact relationship represents an intersection topological relationship between two spatial dimensions: the dimension Highway Segment of type Line and the dimension City of type area.

This works presents and analyses the different components that must be included in a spatial multidimensional model. It defines and presents same examples of models that include spatial dimensions, spatial hierarchies, spatial measures and spatial fact relationship. However, no complete model containing spatial dimensions, spatial measures and a spatial fact relationship is presented, every feature is analysed separately. Finally, the inclusion of temporal feature is done only through the inclusion of the thematic dimension time.

*C. The Model of Miquel and al*

Miquel and al [9] handled the problem of integration of the spatiotemporal data in geospatial data warehouses. They introduced the notion of geometrical spatial dimension as well as the notion of specific or generic dimension. Basing on these dimensions, they proposed two approaches for the modelling of the multidimensional structures for the heterogeneous spatial data presenting problems of integration. These two approaches were applied to an example taken from the domain of the forestry.

The problems of integration handled in the forest inventories are owed to the evolution of the number of attributes, to the change of their semantics from an inventory to an another, to the different definition of the used unit of measure (the notion of units of management) and to the variation of the consequentive domains of the attributes.
The attribute age for example is considered as qualitative in an inventory and quantitative in an other. The following table shows an example of three forest inventories presenting these problems of heterogeneousness.

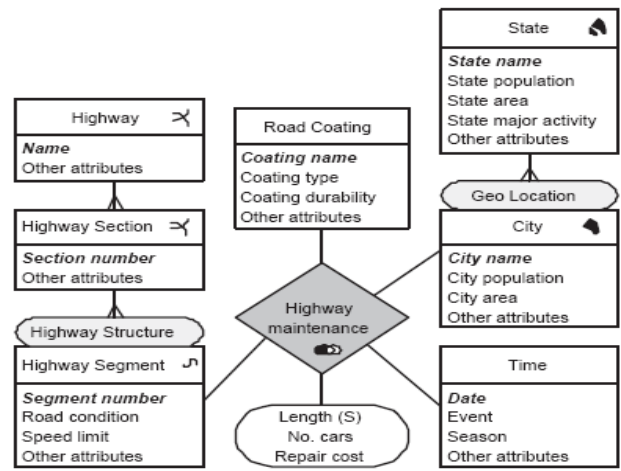

Figure 4: An example of multidimensional schema with spatial elements [6]

The first proposed model is a model in constellation. This model is based on several specific fact tables and a generic fact table. There are so many specific fact tables as considered intervals of time. In this model, every inventory gives place to the constitution of a spatially and temporally specific fact table.

TABLE1 : CHARACTERISTICS OF INVETORIES [9]

| Year | Number of ecoforest Populating | Number of attributes | Unit of management |
|------|-------------------------------|----------------------|--------------------|
| 1973 | 1700 | 5 | Populating Planting |
| 1984 | 2400 | 14 | Unity of landscape Compartment |
| 1992 | 3800 | 19 | ecoforest Populating ecological Polygon forest Station |

The second presented model is a star model. It is based on a single fact table where every measure of the fact table relates to a geometrical cell stemming from the matrix representation of the territory. The associated thematic dimensions are formed by specific members and by generic members.

*D. The model of bedard and al*

Basing on Perceptory, Bedard and al [10] propose a method and a tool for modelling generalisation and multiple representations for a spatial data warehouse. In order to modelise multiple representations, researchers propose three modelling rules to be applied in the object class. These rules are hereunder:

**Rule 1**: "All the generalisations operations are defined by an operation in the object class (in the operation section) whether this process is manuals, semi-automatic or entirely automatic".
**Rule 2**: "All multiple representations are defined using a multiple geometry pictogram in the object class.
**Rule 3**: If the geometry resulting from the generalisation process must to be stored in the system, this geometry will be added in the object class using the PVL multiple geometry pictogram.

In order to model generalisations operations, two major rules are proposed, these rules are hereunder:

**Rule 1**: If the generalisation process is based on a unique operator, the generalisation operation is identified by the name of this operator (Example: Collapse, Simplification, and Displacement).
**Rule 2:** If the generalisation process is made by applying a set of generalisation operators, the generalisation is identified by a general operation called "GEN" and the operators sequence is described in the data dictionary.

In order to better explain the different ideas presented so far, those general modelling rules are applied on an application example used for the design of a spatial data warehouse for the Quebec Ministry of Natural Resources.
Table 2 presents the modelling of multiple representations and generalisations for the waterstream object class.

This example shows that all geometries of the object class Watersetraem could be represented either by a a linear or a polygonal geometry depending on certain geometric rules. In fact, the Waterstream section is presented by a polygon if its area is more than 2000m2 and its minimal width is 150m and it is represented by a line if its area is less than 2000m2 and its minimal width is 20m. So in the object class, we will have either a linear or a polygonal geometry. The geometry of the scale of 20K are generalised to obtain the geometry of 100K and 25OK. The generalisation operation is made by applying a set of generalisation operators, That is why, it is described by the term 'GEN'.The complete set of operators and as well as their order is indicated in the data dictionary

This work presents a solution and a set of rules that allows the modelling of multiple representations and generalisations for spatial data warehouse. However, the integration of this solution in the different components of the multidimensional schema of spatial data warehouse was not explicitly analysed.The inclusion of PVL pictograms facilitates the reading and the comprehension of the schema but it reduces it size significantly, especially when dealing with alternate *geometries.*

TABLE 2: THE MODELLING OF THE OBJECT CLASS:"WATERSTREAM" [10].

| WATERSTREAM |
|---|
| toponym |
| importance |
| |
| GEN 20K → 100K |
| GEN 20K → 100K |
| GEN 20K → 250K |

*E. The Model of Bauzer-Medeiros and al*

Bauzer-Medeiros and al [11] proposed a multidimensional model for the treatment of spatiotemporal data in the field of the road traffic. It is a new domain of application that of the real-time moderate spatiotemporal data.

The paper deals with the integrating of a big volume of spatiotemporal data collected by hundreds of sensors placed on the main highways of a city. These data describe the state of the traffic (circulation) in the form of two measures to know the debit (flow) and the rate of activity (occupation). The debit (flow) expresses the number of vehicles by unit of time while the rate of activity (occupation) translates the relationship between the time of passage of a vehicle and the interval of time of the measure. Other meteorological information (the temperature, the accumulated precipitation) influencing the state of the traffic is taken into account in the conceptual modelling. The multidimensional modelling of a road traffic data warehouse allows users to make decisions with court and long-term. They are decisions going of the macro scale (such as the construction of a freeway) until the punctual decisions (such as the adjustment of a traffic light).

In order to achieve these objectives, the following multidimensional model is presented. It is a star schema composed of five dimensions and two measure.

1. The dimension *Day*: this dimension serves for analysing relations between the moderate values (debit and flow of circulation) and the temporal activities (such as strikes, parades, festivals, sportive events).
2. The dimension *Schedules*: this dimension allows analyzing the road traffic according to The clock or according to the rhythm of the human activity at several levels of granularity. (Example: The morning, the afternoon, the evening).
3. The dimension *Sensor*: It allows realizing studies on the state of the traffic in every place. The data stemming from this dimension can be seen at several levels of aggregation such as the topology of the road graph, the geography of the city, the influences that can have specific urban elements situated near
4. The dimension *weather*: it allows making analyses between the state of the traffic and the meteorological data.
5. The dimension *annotations*: it reflects elements of the events which can influence the state of the traffic such as a breakdown of traffic lights, the snowfall …

The proposed model allows the analyse the state of the traffic (debit) and the rate of activity (occupation) in a given day, during a fixed schedule, by a given sensor, for certain weather conditions and for a given event. The spatial aspect is represented by a non geometrical spatial dimension (dimension sensor) while the temporal aspect is represented by the hourly dimension and days. However, the model does not include spatial measures. Through these two dimensions, this model deals with the evolution of the data in space and time.

*F. The model of Glorio and al*

Glorio and al [12] propose to introduce spatiality in the classical multidimensional models. They propose a set of extensions that concerns spatial levels and spatial measures. Indeed, they introduce new stereotypes named *SpatialLevel* () and *SpatialMeasure* (). These stereotypes are endowed by a set of properties to describe their geometric shapes. All these extensions are implemented based on the UML profile. The researchers also propose a set of rules for the transition of the model conceptual to the logical model.

*G. The model of Bâazaoui Zghal and al*

Bâazaoui Zghal and al [13] propose a metamodel for building spacial data marts (spatial datamart). the proposed spatial datamart is identified by a unique identifier that does not change during its life cycle. It is characterized by a set of attributes such as name, type of activity and the type of analysis. Each data mart is characterized by a multidimensional schema composed by one or more tables of facts, dimensions and measures. The fact table is the subject of analysis and is presented in this model by a class an object of a set of attributes. in this proposal, dimensions and measures may be spatial or non spatial. A dimension or a measure could be visualized as a non-spatial level of the hierarchy and then it becomes a spatial level in the case of generalizations. Fig 5 shows the proposed metamodel.

|  | **Model of Malinowski and Zimany** | **Model of Miquel and al** | **Model of Bauzer-Medeiros and al** | **Model of Stefanovic and al** | **Model of Glorio and al** | **Model of Bâazaoui Zghal and al** |
|---|---|---|---|---|---|---|
| **Explicit hierarchies in dimensions** | × | No hierarchies in the model | No hierarchies in the model | No hierarchies in the model | - | × |
| **Symmetric treatment of dimensions and measures** | × | - | - | × | - | - |
| **Multiple hierarchies in each dimension** | - | No hierarchies in the model | No hierarchies in the model | No hierarchies in the model | - | - |
| **Support for correct aggregation** | - | No hierarchies in the model | No hierarchies in the model | No hierarchies in the model | × | × |
| **Non strict hierarchies** | × | No hierarchies in the model | No hierarchies in the model | No hierarchies in the model | - | - |
| **Many to many relationships between facts and dimensions** | × | × | × | × | - | × |
| **Manipulate change and time** | × | × | × | × | - | × |
| **Manage several levels of granularity** | - | No hierarchies in the model | No hierarchies in the model | No hierarchies in the model | × | × |

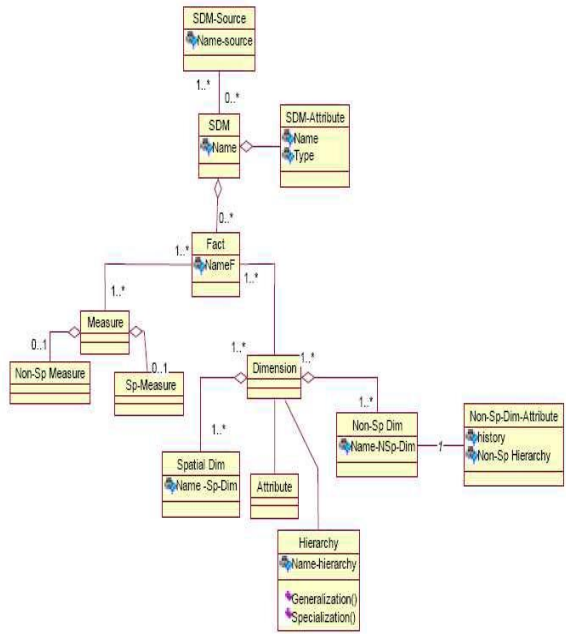

Figure 5: A métamodel for the construction of a spatial data

## III. COMPARATIVE STUDY OF SPATIAL AND SPATIO-TEMPORAL MULTIDIMENSIONAL MODELS

The following section presents a comparative study that provide a general, comparative view of the spatial and spatio-temporels multidimensional models that have been presented and discussed. The different models are evaluated against these six requirements inspired from [14, 15]

1) **Explicit hierarchies in dimensions**: This requirement analyse the way in which the hierarchies are organised in the model. In fact in a multidimensional schema, hierarchies must be captured explicitly so that the relation between the different levels in the dimension will be available.

2) **Symmetric treatment of dimensions and measures**: The model should allow a symmetric treatment of dimensions and measures. This means that a dimension can be presented as a measure and also a measure can be treated as a dimension.
   **Example**: The attribute age for patients can be treated as a measure allowing analysis based on computation. It can also be modelled as a dimension which allows to group patients into age groups.

3) **Multiple hierarchies in each dimension**: to aggregate data, it can be more than one path in one dimension. Exemple: suppose that we have the dimension Time. Day's rolls up to weeks and to months, but weeks do not roll up to months. To model this, multiple hierarchies in each dimension are needed.

4) **Support for correct aggregation**: This requirement means that the model should allow getting "correct" results when aggregation data. In fact, the result of aggregations operations should have a meaningful explanation. Avoiding double counting is one aspect on this requirement.
   Example: Suppose that we have the following hierarchies: Diagnostic → Family diagnostic → Diagnostic group. If a user asks a query is order to get the number of patients in every diagnostic group, the same patient should be counted once per group, despite that the patient has several diagnostics in a group.

5) **Non strict hierarchies:** The hierarchies in a dimension are not always strict. The model should be able to handle non strict hierarchies.

6) **Many too many relationships between facts and dimensions**: the model should support not only the classical many-to-one relationship between facts and dimensions. It should also support many-to-many relationship.

We notice that all the presented models allow an explicit representation of the hierarchies at the level of the dimensions except for the model of Miquel and al [9]. On the other hand, none of the proposed models allows an explicit representation of the dimensions and measures. Also, no model supports a multiple relationships between the dimensions and the facts.

The model of Malinowski and Zimany [6] calls upon the idea of the symmetric treatment of the dimensions and the measures. It allows verifying this condition not for all the dimensions of model but only for the dimension Location treated as measure and as dimension. Concerning the manipulation of the change and of time, Malinowski and Zimany mention the necessity of managing the evolution of the spatial data and they propose the dimension time as a

Table3: Comaprative study betwen spatial and spatio-temporal multidimensonnal models means to represent the evolution of these data. On the other hand, the model is not interested either in the problem of the correct aggregations or in the multiple hierarchies. However, it allows the sharing of the hierarchies between several dimensions.

Concerning the model of Miquel and al [9], it verifies none of the presented conditions, exception made for the manipulation of the change and of time. Also, the model of Bauzer-Medeiros [11] and al verifies none of the conditions except the management of the change. In fact, it is interested in the modelling of the evolutions which can affect the transport network according to schedules, place, topology of the zone, mode of the activities. Thus, these two models are rather interested in the modelling and the integration of the variation of the spatiotemporal data.

The work of Bédard and al [10] was not includes in this comparative study. In fact, it does not propose a multidimensional model, but rather a set of rules for the modelling of the multiple representations in the data warehouse.

## IV. CONCLUSION

We notice that the majority of the multidimensional models integrate spatiality in the conceptual multidimensional modelling basing on non geometrical spatial dimensions. The geometrical and mixed spatial dimensions are neglected in the multidimensional modelling. The elaboration of multidimensional models which are interested in the modelling and the analysis of this type of measure is a research subject susceptible to be better investigated

In a multidimensional model including more than a spatial dimension, to define a topological relation between the geometrical or mixed spatial dimensions is often necessary. It makes important to define spatial relations between the geometrical or mixed spatial dimensions.

In spite that some models [6, 9] are interested in the multidimensional modelling of the temporal data but this modelling remains independent from the spatial aspect. The proposal of a spatiotemporal multidimensional models including dimensions and spatiotemporal measures allows to facilitate the analysis of data and to integrate aspects which were neglected.

The subject of the evolution of the data is treated according to diverse aspects. Indeed, certain models are interested in the modelling of the evolution of semantics of the data and their values [9].Others are interested in the modelling of the data further to the change of the scales of measures [10]. Others treat particular aspects of the change of the data which are connected to the nature of the studied phenomenon such as the modelling of the evolution of the state of the traffic [11].

Because the data used for decision-making change and evolve in a continuous way and because the nature of change depends on the nature of the application. The consideration of the evolution and the change of the data in the multidimensional modelling turns out an interesting subject which deserves to be better investigated according to the other aspects and in adequacy with the nature of the studied domains.

In conclusion, we can say that the exploitation of phases of analysis and conception to choose better "what" to include in the warehouses of data and "how" to include has always been a fields of research which deserves to be better exploited.